\documentstyle[11pt,epsfig,aaspp4]{article}
\tighten 
\begin{document}

\baselineskip 5.0ex

\newcommand{\lya}{Lyman~$\alpha$}
\newcommand{\lyb}{Lyman~$\beta$}
\newcommand{\za}{$z_{\rm abs}$}
\newcommand{\ze}{$z_{\rm em}$}
\newcommand{\cmtwo}{cm$^{-2}$}
\newcommand{\nhi}{$N$(H$^0$)}
\newcommand{\nzn}{$N$(Zn$^+$)}
\newcommand{\ncr}{$N$(Cr$^+$)}
\newcommand{\degpoint}{\mbox{$^\circ\mskip-7.0mu.\,$}}
\newcommand{\halpha}{\mbox{H$\alpha$}}
\newcommand{\hbeta}{\mbox{H$\beta$}}
\newcommand{\hgamma}{\mbox{H$\gamma$}}
\newcommand{\kms}{\,km~s$^{-1}$}      
\newcommand{\minpoint}{\mbox{$'\mskip-4.7mu.\mskip0.8mu$}}
\newcommand{\mv}{\mbox{$m_{_V}$}}
\newcommand{\Mv}{\mbox{$M_{_V}$}}
\newcommand{\peryr}{\mbox{$\>\rm yr^{-1}$}}
\newcommand{\secpoint}{\mbox{$''\mskip-7.6mu.\,$}}
\newcommand{\sqdeg}{\mbox{${\rm deg}^2$}}
\newcommand{\squig}{\sim\!\!}
\newcommand{\subsun}{\mbox{$_{\twelvesy\odot}$}}
\newcommand{\et}{et al.~}

\def\ltsima{$\; \buildrel < \over \sim \;$}
\def\simlt{\lower.5ex\hbox{\ltsima}}
\def\gtsima{$\; \buildrel > \over \sim \;$}
\def\simgt{\lower.5ex\hbox{\gtsima}}
\def\arcs{$''~$}
\def\arcm{$'~$}
\vspace*{0.1cm}
\title{METAL ABUNDANCES AT $z < 1.5$: FRESH CLUES TO THE CHEMICAL
ENRICHMENT HISTORY OF DAMPED Ly$\alpha$ SYSTEMS\altaffilmark{1}}

\vspace{1cm}
\author{\sc Max Pettini}
\affil{Royal Greenwich Observatory, Madingley Road, Cambridge, CB3 0EZ, UK}
\author{\sc Sara L. Ellison}
\affil{Institute of Astronomy, Madingley Road, Cambridge, CB3 0HA, UK}
\author{\sc Charles C. Steidel\altaffilmark{2}}
\affil{Palomar Observatory, Caltech 105--24, Pasadena, CA 91125}
\author{\sc David V. Bowen}
\affil{Royal Observatory, Blackford Hill, Edinburgh, EH9 3HJ, UK}

\altaffiltext{1}{Most of the data presented herein were obtained at the 
W.M. Keck Observatory, which is operated as a scientific partnership 
among the California Institute of Technology, the University of 
California, and the National Aeronautics and Space Administration. The 
Observatory was made possible by the generous financial support of the 
W.M. Keck Foundation.}
\altaffiltext{2}{NSF Young Investigator}

\newpage
\begin{abstract}
We explore the redshift evolution of the metal content of damped \lya\ 
systems (DLAs) with new observations of four absorbers at $z < 1.5$; together 
with other recently published data, there is now a sample of ten systems
at intermediate redshifts for which the abundance of Zn has been 
measured. The main conclusion is that the column density--weighted 
mean metallicity, 
${\rm [} \langle{\rm Zn/H}\rangle {\rm ]} = -1.03 \pm 0.23$ (on a 
logarithmic scale), is not 
significantly higher at $z < 1.5$ than at earlier epochs, 
despite the fact that the comoving star formation rate density of 
the universe was near its maximum value at this redshift.
Gas of high column density and low metallicity dominates the 
statistics of present samples of DLAs at all redshifts.

For three of the four DLAs our observations include absorption lines of 
Si, Mn, Cr, Fe, and Ni, as well as Zn. We argue that
the relative abundances of these elements are consistent with a moderate 
degree of dust depletion which, once accounted for, leaves no room for 
the enhancement of the $\alpha$ elements over iron seen in metal poor 
stars in the Milky Way. This is contrary to previous assertions that DLAs have 
been enriched solely by Type II supernovae, but can be understood if the 
rate of star formation in the systems studied proceeded more slowly than 
in the early history of our Galaxy. 

These results add to a growing body of data all pointing to the conclusion 
that known DLAs do not trace the galaxy population 
responsible for the bulk of star formation. Possible reasons are
that sight-lines through metal rich gas are systematically 
underrepresented because the background QSOs are reddened, 
and that the most actively star forming galaxies are also the most
compact, presenting too small a cross-section to have been probed yet with 
the limited statistics of current samples.
\end{abstract}

\newpage
\section{INTRODUCTION}

Damped \lya\ systems have been studied extensively at redshifts
greater than $z = 1.5$ to determine the consumption of interstellar
gas into stars (Lanzetta, Wolfe, \& Turnshek 1995; Storrie-Lombardi,
McMahon, \& Irwin 1996); measure the abundances of metals and dust 
(Fall \& Pei 1993; Lu et al. 1996; Pettini et al. 1997a,b); explore the
kinematics of forming galaxies (Prochaska \& Wolfe 1997; Haehnelt,
Steinmetz, \& Rauch 1998); and probe the spectrum of primordial
density fluctuations on galactic scales (Gardner et al. 1997; 
Peacock et al. 1998).
However, somewhat paradoxically, until recently it has proved
difficult to extend these studies to lower redshifts where in
principle it should be easier to establish the connection between
damped systems and galaxies in the Hubble sequence. The reason is that
ultraviolet observations are required to identify a damped \lya\ line
at $z < 1.5$; the archive of {\it Hubble Space Telescope} QSO spectra
has taken several years to grow to a size sufficient for assembling
even a modest sample of DLAs. 

The picture which is emerging is far from clear yet. Estimates of the
number density of damped systems per unit redshift at $z < 1.5$ range
from $dN/dz \simeq 0.1$ (Turnshek 1998) to $dN/dz \simeq  0.02$
(Jannuzi et al. 1998); with the present uncertainties it is hard to
discern any significant 
redshift evolution in the cosmological mass density
$\Omega_{\rm DLA}$. Imaging searches for the galaxies responsible for
producing DLAs, from space and from the ground, have revealed a highly
diverse population of absorbers which so far includes low surface
brightness galaxies, dwarfs, and even one early-type galaxy, as well
as spirals (Steidel et al. 1994; Le Brun et al. 1997; Lanzetta et al.
1997; Rao \& Turnshek 1998). This is at odds with the results of local
field surveys of H~I in 21 cm emission which show that large spiral
galaxies are the major contributors to the local H~I mass function, at
least down to $M_{\rm H~I} > 10^8 M_{\odot}$ (see, for example, the
comprehensive discussion by Zwaan 1998).

In this paper we consider the chemical evolution of DLAs at $z < 1.5$.
The compilation of Zn and Cr abundances by Pettini et al. (1997a)
included only four measurements in this redshift interval. Here we
present observations of four new systems which, together with recently
published data for two others, bring the total to ten and allow us to
follow the chemical enrichment of DLAs down to $z = 0.4$\,.
Furthermore, for three of the new cases our spectra include lines of
several other species, as well as Zn~II and Cr~II, and we look to the
pattern of relative element abundances for clues to the chemical
history of the gas.

The paper is arranged as follows. The observations and data reduction
are described in \S2, while \S3 deals with the derivation of column  
densities and element abundances. In \S4 we use  
the enlarged data set to assess whether there is any redshift evolution 
in the metallicity of DLAs at $z < 1.5$. Our analysis of 
the abundance ratios is presented in \S5; finally in \S6 we discuss 
the implications of our results for the interpretation
of damped \lya\ systems and emphasize the relevance of this work 
for recent ideas on the nucleosynthetic origin of 
some elements. \\

\section{OBSERVATIONS}
\subsection{{\it HST} Data}

Three of the four DLAs considered here (Q1247+267, Q1351+318, and
Q1354+258) were identified in a trawl of the {\it HST} FOS data
archive; details of the original observations are given in Table 1. 
The pipeline calibrated spectra were resampled to a linear dispersion
of 0.51~\AA\ per pixel (one quarter diode steps); in Q1247+267 and
Q1351+318 small corrections for
scattered light ($\approx 3-4$\%) were found to be necessary to bring
the cores of the damped \lya\ lines to zero flux. 
Figure 1 shows portions of the spectra, normalised to the
underlying QSO continua, together with our best fits to the damped
profiles. Q1247+267 is a bright QSO ($V = 15.8$) and the 4,500~s FOS
exposure produced a spectrum of moderately high signal-to-noise ratio,
S/N = 28. Q1351+318 and Q1354+258 are fainter and were observed for
shorter exposure times giving lower quality spectra. However, as can
be seen from the last column of Table 1, the column density of neutral
hydrogen can be deduced with an accuracy of better than 25\% in all
three cases and ranges from \nhi\ $= 7.5 \times
10^{19}$~cm$^{-2}$ in Q1247+267 to $3.5 \times 10^{21}$~cm$^{-2}$ in
Q1354+258.

The fourth QSO is the gravitationally lensed pair Q0957+561A and B. It
shows an absorption system at $z_{\rm abs} = 1.3911$, near the
emission redshift $z_{\rm em} = 1.4136$, the damped nature of which
was first realized by Turnshek \& Bohlin (1993) from {\it IUE}
archival observations and subsequently confirmed with {\it HST} FOS
spectra by Michalitsianos et al. (1997) and Zuo et al. (1997). Both
sight-lines through the $z_{\rm abs} = 1.3911$ absorber (the
separation is $0.24~h_{70}^{-1}$~kpc for $q_0 = 0.1$) intersect gas
with large values of \nhi ; in Table 1 we quote the values deduced by
Zuo et al. whose spectra have the higher S/N.

We note that three of the four DLAs studied here have values
of neutral hydrogen column density  which are lower than the threshold
\nhi\ $= 2 \times 10^{20}$~cm$^{-2}$
originally adopted by Wolfe et al. (1986), although only in one case 
(the \za = 1.22319 system in Q1247+267) significantly so.
This reflects the shift of the column density distribution
toward lower values of \nhi\ at $z < 1.5$ first found by Lanzetta
et al. (1995).\\

\subsection{Optical Data}

The optical spectra of the first three QSOs in Table 1 were
recorded at high spectral resolution with the HIRES echelle
spectrograph (Vogt et al. 1994)
on the Keck I telescope on Mauna Kea, Hawaii in
February -- March 1998. Relevant details of the observations are
collected in Table 1. In 0.6--0.7~arcsec seeing we used a
0.86~arcsec wide entrance slit which projects to 3 pixels on the
2048x2048 Tektronix CCD detector, giving a resolution of 6
km~s$^{-1}$~FWHM. The echelle and cross-disperser angles were
adjusted so as to record all lines of interest 
in the three DLAs between approximately
3800 and 6200~\AA; in the spectra
of Q1247+267 and Q1351+318 we covered all prominent  absorption
lines from Si~II~$\lambda 1808$ to the Mg~II~$\lambda\lambda
2796,2803$ doublet, while in Q1354+258 all lines between
Fe~II~$\lambda 1608$ to Mn~II~$\lambda 2594$ were included.

The echelle spectra were extracted with Tom Barlow's customised
software package, wavelength calibrated by reference to the spectra of
a Th-Ar hollow cathode lamp, mapped onto a linear wavelength scale,
and divided by a smooth continuum. The rms deviations from the
continuum fit give a measure of the final S/N of the data. Since the 
S/N generally decreases with decreasing wavelength, reflecting the lower
efficiency of HIRES in the blue, we have listed in column (10) of
Table 1 indicative values which apply to most absorption lines in each
DLA. 

In Figures 2, 3, and 4 we have reproduced examples of absorption lines
of varying strengths in each damped system. Again it can be seen that
the spectrum of the bright QSO Q1247+267 is of particularly high
precision, but in all three cases the high resolution of the echelle
data results in very sensitive detection limits of only a few m\AA\ in
rest frame equivalent width $W_{0}$ (column (12) of Table 1). Tables
2, 3 and 4 list the absorption lines detected in each DLA; resolved
components within a complex absorption line are indicated by a lower
case letter. The errors quoted for $W_{0}$ reflect the counting
statistics only and do not take into account uncertainties in the
continuum placement nor in the wavelength interval over which the
equivalent width summation is carried out. Although difficult to
estimate, the latter is probably the major source of error affecting our
values of $W_{0}$.

The spectra of Q0957+561A and B were recorded in March 1997 at
intermediate resolution with the cassegrain spectrograph of the
William Herschel Telescope (WHT) on La Palma, Canary Islands, set to
cover the Zn~II~$\lambda\lambda2026, 2062$ and
Cr~II~$\lambda\lambda2056, 2062, 2066$ multiplets at \za\ = 1.3911\,.
The acquisition and reduction of the data followed the procedures
described by Pettini et al. (1997a). As can be seen from Figure 5, no
absorption lines were detected to a limiting equivalent width
$W_0$($3\sigma$) $\simeq 30$~m\AA.\\

\section{ION COLUMN DENSITIES AND ELEMENT ABUNDANCES}

As can be seen from Figures 2, 3 and 4, the profiles of the
absorption lines in the three DLAs recorded with HIRES are
complex, indicating the presence of multiple absorbing clouds
along each sight-line. In order to measure ion column densities,
we have used the VPFIT package written by Bob Carswell to
decompose the absorptions into individual components. For each
component VPFIT returns the values of redshift, velocity
dispersion parameter {\it b} ($b = \sqrt{2} \sigma$, where
$\sigma$ is the one-dimensional velocity dispersion of the
ions along the line of sight, assumed to be Gaussian),
and ion column density {\it N} which best fit the observed
absorption; the number of components was kept to the
minimum required to make the differences between calculated and
observed profiles consistent with the S/N of the data.
We adopted the compilation of wavelengths and $f$-values by
Morton (1991) with the revisions proposed by Savage \& Sembach (1996).

In each DLA the profile decomposition into multiple components is
determined by the Fe~II lines. Our observations cover seven
different lines from Fe~II UV multiplets spanning a wide range of
$f$-values, from $f = 0.00182$ for Fe~II~$\lambda 2249$ to $f =
0.3006$ for Fe~II~$\lambda 2382$. Components of progressively
lower column density are revealed as we move up this sequence,
particularly in Q1351+318 (see right hand panels in Figure 3).
Details of the profile fits to the Fe~II lines are collected in
Tables 5, 6, and 7 and an example is reproduced in Figure 6. The
values of $b$ and $z_{\rm abs}$ determined from the Fe~II lines
were found to fit well the absorption lines of all the other 
first ions (reduced $\chi^2 \simlt 1.1$), 
leaving the column density as the only free parameter.

By adding the contributions of different absorption components we
deduced the total column densities listed in columns (4) to (9)
of Table 8. It is important to realise that the total values of
$N$ do {\it not} depend on the details of the model fits because
for each species our HIRES spectra include weak absorption lines
which lie on the linear part of the curve of growth. Accordingly,
we have not included Mg$^+$ in Table 8 because the
Mg~II$\lambda\lambda 2796, 2803$ doublet lines {\it are}
saturated in all three DLAs and the corresponding column
densities cannot be determined reliably, even though the profiles
can be fitted satisfactorily. The error estimates assigned to the
values of $N$ in Table 8 reflect the uncertainties in the
equivalent widths and the agreement between different absorption
lines of the same ion. 

Assuming that for the elements observed the first ions are the
dominant ionization stages in the H~I gas producing the damped
\lya\ lines (their ionization potentials are higher than that of
neutral hydrogen), the total element abundances can be deduced
directly by dividing the values of {\it N} in columns (4) to (9)
by the values of \nhi\ in column (3) of Table 8. Comparison with
the solar abundance scale of Anders \& Grevesse (1989) finally
gives the relative abundances listed in Table 9. If some of the
first ions absorption arises in H~II gas, the derived abundances
are upper limits to the true abundances in the DLAs. 
However, previous detailed analyses of this point
(e.g. Viegas 1995; Prochaska \& Wolfe 1996) have generally concluded that 
even at the relatively low values of \nhi\ of some of the DLAs considered 
here such ionization corrections are likely to be small.
This conclusion is reinforced by the decreasing intensity of the
ionizing background at $z < 1.5$ (Kulkarni \& Fall 1993).  
We now briefly describe each system in turn.\\

{\it Q1247+267; $z_{\rm abs}$ = 1.22319}.  The Fe~II lines in this
system show absorption from two components separated by 13.5
\kms\ with the higher redshift component, at \za\ = 1.223202, 
contributing 95\% of the total column density (Table 5). The
strongest lines in this DLA, Mg~II$\lambda\lambda 2796, 2803$,
show additional weak absorption at velocities $v_{\rm rel} \simeq
-45$ and $+80$~\kms\ relative to the main component (Figure 2).
The Zn~II and Cr~II lines are among the weakest detected to date,
with rest-frame equivalent widths $W_0 = 5 - 7$~m\AA, implying
abundances of less than 1/10 of solar (Table 9).

{\it Q1351+318; $z_{\rm abs}$ = 1.14913}.  This is a complex
system with absorption spanning $\sim 400$~\kms. The line
profiles are reminiscent of those seen towards stars in the
Magellanic Clouds (e.g. Blades et al. 1988) and towards some
supernovae in nearby galaxies (e.g. Bowen et al. 1994),
suggesting that the sight-line to Q1351+318 may intersect two
companion galaxies near $z = 1.1491$. The Fe~II lines require a
minimum of 13 components for a satisfactory fit (Table 6 and
Figure 6); additional components can be discerned in Mg~II
(Figure 3). Despite this complexity, 77\% of the total column
density of Fe~II is due to only two components at \za\ = 1.149033
and 1.149139 (respectively numbers 2 and 3 in Table 6 and in
Figure 6); the weak lines reproduced in the left-hand panels of
Figure 3 arise mostly in these two `clouds'. 

We deduce an abundance of Zn of $\sim 1/2$ solar (Table 9). Thus
we have found a second example of a DLA system with abundances
consistent with the metal enrichment history of the Milky Way
stellar disk which, near the Sun, had a mean [Fe/H] $\approx -0.4$
at $z = 1.1$ (see Figure 14 of Edvardsson et al.
1993)\footnote{We use the conventional notation where [X/H] = log
[$N$(X)/$N$(H)]$_{\rm DLA}$ $-$ log [$N$(X)/$N$(H)]$_{\odot}$.}
The other example is the \za\ = 1.0093 absorber in
EX~0302$-$223 studied by Pettini \& Bowen (1997).

{\it Q1354+258; $z_{\rm abs}$ = 1.42004}.  This is a simple
absorption system with one component, at \za\ = 1.420053
contributing 97\% of the total column density. The Zn~II
absorption lines are weak, despite the large neutral hydrogen
column density, \nhi\ = $3.5 \times 10^{21}$~cm$^{-2}$; we deduce 
a Zn abundance of $\sim 1/40$ solar.

{\it Q0957+561A, B; $z_{\rm abs}$ = 1.3911}.  The Zn and Cr lines are
below our detection limit along both sight-lines. The more
stringent limits are those for Q0957+561A where \nhi\ is higher;
we find that Zn and Cr are less abundant than $1/6$ and $1/13$
solar respectively.

Finally, in the discussion below we include measurements of the
Zn abundance in two other intermediate redshifts DLAs which have
become available since the compilation by Pettini et al. (1997a).
In their {\it HST} FOS study Boiss\'{e} et al. (1998) concluded
that [Zn/H] = $-0.47 \pm 0.15$ in the \za\ = 0.3950 DLA towards
PKS~1229$-$021, where they measured log~\nhi\ = 20.75 $\pm
0.07$~cm$^{-2}$. de la Varga \& Reimers (1998) reported [Zn/H] =
$-1.46$ at \za\ = 0.68 in HE~1122$-$168 from ground-based echelle
spectra, having established that this absorption system is damped
on the basis of {\it HST} FOS observations (log~\nhi\ = 20.45
$\pm 0.05$~cm$^{-2}$).\\

\section{REDSHIFT EVOLUTION OF THE METALLICITY OF DAMPED Ly$\alpha$ SYSTEMS}

Figure 7 shows the full set of measurements of the abundance of Zn 
in 40 DLAs from \za\ = 0.3950 to 3.3901. The main conclusion is that
even though the number of measurements at \za\ $< 1.5$ has increased
from four to ten, the overall picture has not changed 
from that which could be gleaned from the survey of Pettini et al. (1997a).
Evidently, the metallicity of damped \lya\ systems 
does not increase  with 
decreasing redshift, as may have been 
expected if they traced the bulk of 
the galaxy population in an unbiased way.
Qualitatively, Figure 7 suggests some mild evolution in that, 
if we treat the upper limits as detections, six out of nine DLAs
at \za\ $< 1.5$ have abundances greater than 1/10 solar,
whereas at \za\ $> 1.5$ only ten out of thirty
are this metal-rich.
Quantitatively, however, we are interested in the column density--weighted 
metallicity:

\begin{equation}
	{\rm [} \langle{\rm Zn/H}_{\rm DLA}\rangle {\rm ]} = 
        {\rm log}\langle{\rm(Zn/H)}_{\rm DLA}\rangle-{\rm log~(Zn/H)}_{\odot}, 
	\label{}
\end{equation}
\noindent where
\begin{equation}
	\langle{\rm (Zn/H)}_{\rm DLA}\rangle = 
        \frac{\sum\limits_{i=1}^{n} N{\rm(Zn}^+{\rm)}_i}
        {\sum\limits_{i=1}^{n} N{\rm(H}^0{\rm)}_i} , 
	\label{}
\end{equation}

\noindent which is a measure of the degree of metal enrichment of the 
population as a whole.
Values of 
${\rm [} \langle{\rm Zn/H}_{\rm DLA}\rangle {\rm ]}$
in different redshift intervals
are listed in Table 10 and plotted in Figure 8.
Again we have treated the upper limits as if 
they were detections but, as shown by Pettini et al. (1997a),
this assumption does not affect significantly 
the accuracy of our estimates 
of ${\rm [} \langle{\rm Zn/H}_{\rm DLA}\rangle {\rm ]}$ because 
the upper limits are all from low column density systems which make 
small contributions to the summations in equation (2). 
The errors quoted were derived with the bootstrap method
(Efron \& Tibshirani 1993), using 500 random samples of the data
to form a distribution 
of values of ${\rm [} \langle{\rm Zn/H}_{\rm DLA}\rangle {\rm ]}$
from which the standard deviation could be estimated.\footnote{The 
errors so derived are smaller, 
and more realistic,
than those given in Pettini et al. 
(1997a) which were simply the standard deviation
of {\it individual} values of 
[Zn/H] from the column density-weighted mean.}  

It can be seen from Figure 8 that the metal content of the DLA population 
does not increase at $z < 1.5$\,. Although the frequency of 
metal-rich absorbers may be higher, {\it the census of metals at all 
redshifts is dominated by high column density systems of low metallicity}.
This result contrasts with the redshift evolution of the comoving star 
formation rate density which is near its maximum value 
at $z = 1 - 2$ (Madau, Pozzetti, \& Dickinson 1998).

A plausible explanation is that present compilations of 
DLAs are biased against metal rich, high column density systems. 
Such systems may be intrinsically rare (and may therefore require larger 
samples of QSO sight-lines to be intersected) 
because in the most metal-rich galaxies much of the gas 
has been turned into stars (Wolfe \& Prochaska 1998), 
or may be missed because
the associated dust extinction preferentially removes QSOs 
in these directions from magnitude limited samples, as reasoned by 
Pei \& Fall (1995). 
This second selection effect is likely to be particularly 
severe at $z < 1.5$, simply because 
the limited aperture of {\it HST}--- required for identifying 
a DLA at $z< 1.5$---imposes a brighter magnitude 
limit than is the case for ground-based surveys. 
It remains to be seen how important this bias is at high 
redshift. All that can be said at the moment is that the only indication 
of a redshift evolution in the metallicity of DLAs is the increase 
between $z = 4$ and 3 suggested by the data in Figure 8 and confirmed by 
the [Fe/H] measurements of Lu, Sargent, \& Barlow (1998).
We return to this point in the Discussion at \S6 below.\\

\section{ELEMENT RATIOS}

The pattern of element abundances in the three DLAs observed with HIRES
is reproduced in Figure 9. Also shown in the figure are the relative 
abundances of the same elements in interstellar 
clouds in the halo of our Galaxy, as compiled by Savage \& 
Sembach (1996) from {\it HST} observations of stars
a few kpc from the Galactic plane. 
Here the missing fractions of Si, Mn, Cr, Fe, and Ni, relative to Zn,
are thought to reflect the degree to which these elements have been  
removed from the gas-phase and incorporated into dust particles;
on the other hand, Zn (and S which is not shown here) are undepleted
and show essentially solar abundances in the gas.
We have chosen halo (as opposed to disk) clouds 
for the comparison because this seems to be the regime in the local 
interstellar medium (ISM) which most closely resembles the mild dust 
depletions typical of DLAs (Pettini et al. 1997b; Welty et al. 1997). 

In order to interpret the element ratios seen at high redshift, it is 
necessary to 
distinguish the effects of dust depletion 
from inherent departures from solar relative 
abundances which, if present, would offer clues to previous history of 
star formation in the galaxies associated with the DLAs.
There are two complications here. 
First, many of our abundance measurements have 
been derived from weak transitions whose $f$-values have undergone 
significant revisions in recent years and may  
still be somewhat uncertain (see Table 2 of Savage \& Sembach 1996).
Second, and potentially more important, we do not know if the depletion 
pattern of halo clouds also applies to the ISM of high redshift galaxies, 
where different physical conditions---such as lower metallicities, higher 
equilibrium temperatures, different rates of supernova induced 
shocks---may all have a bearing on the composition of dust.
Therefore, the two effects are best disentangled at low abundances
where we suspect that dust depletions may be reduced 
(Pettini et al. 1997b) and intrinsic deviations from solar ratios are 
expected to be more pronounced. In the present  the 
abundance measurements  
towards Q1354+258 and Q1247+267 ([Zn/H] $= - 1.61$ and $-1.05$ 
respectively) may be most instructive.

\subsection{Chromium, Iron, and Nickel}

Among the elements covered these are the ones which are 
most readily incorporated into dust; we therefore consider them first 
in order to assess the levels of dust depletions.
In Galactic stars of metallicity [Fe/H] $\simgt -2$ 
all three elements track each other (and Zn) closely (McWilliam 1997 and 
reference therein). In all three DLAs studied Cr, Fe, and Ni 
are less abundant than Zn (Table 9) by relative factors which 
are similar to those seen in halo clouds; 
thus Fe is somewhat more depleted than Cr 
and Ni is more depleted than Fe (by even larger factors than
in local halo clouds). 
The Cr and Fe abundances are so similar ([Cr/Fe] $\simlt -0.2$) 
over a range of depletions 
that one may reasonably question whether the difference is due to 
inaccuracies in the $f$-values---the same transitions have been used
for abundance measurements in DLAs and in halo clouds and, despite
the numerous Fe~II lines available, $N$(Fe$^+$) is essentially fixed by 
the two weakest lines at $\lambda 2249.8768$ and $\lambda 2260.7805$\,. 
On the other hand, the enhanced depletion of Ni is probably 
too large to be attributed entirely to such uncertainties.
As there is no evidence in stars that [Ni/Fe] deviates by more 
than $\pm 0.1$ dex over the full interval 
$-4 \leq {\rm [Fe/H]} \leq 0$ (McWilliam et al. 1995; Ryan, Norris, \& 
Beers 1996),
the most plausible interpretation is that most of the Ni is in solid form 
in all three DLAs, as in halo clouds. Thus, even in 
the \za\ = 1.42004 DLA in Q1354+258, where the overall metallicity is low
([Zn/H] = $-1.61$) and the depletions of other elements 
are not severe ([Cr/Zn] = $-0.20$ and [Fe/Zn] = $-0.42$), 
it appears that 
less than 10\% of the Ni remains in the gas ([Ni/Zn] = $-1.07$).

The finding that the relative abundances of Cr, Fe, and Ni are 
similar to those produced by grain depletion 
supports the interpretation of their low abundances relative to Zn as 
being due to the presence of dust, rather than to an `anomalously high' 
abundance of Zn in DLAs (for which there is no other observational basis),
as speculated by Lu et al. (1996) 
and McWilliam (1997).

\subsection{Manganese}

It can be seen from Figure 9 and Table 9 that Mn is consistently less 
abundant than Cr and Fe, even though these three elements are normally 
depleted by similar amounts in the ISM. The most straightforward 
explanation is that in DLAs, as in Galactic stars, the {\it intrinsic} 
abundance of Mn decreases with decreasing metallicity.
If we assume that Cr and Mn are depleted by similar amounts 
(and therefore adopt ($1 - 10^{\rm [Cr/Zn]}$) 
as the fraction of Mn in solid form)
we deduce intrinsic underabundances 
[Mn/Zn] = $-0.23$, $-0.36$, and $-0.51$ in 
Q1351+318 ([Zn/H] = $-0.36$), Q1247+267 ([Zn/H] = $-1.05$),
and Q1354+258 ([Zn/H] = $-1.61$) 
respectively.\footnote{Higher intrinsic Mn abundances, by a factor of
$\sim 0.15$~dex, are obtained if Mn is depleted like Fe, rather than Cr.}
These values are in good agreement with 
those measured in stars, if Zn is taken as a proxy for Fe
(see Figure 12 of McWilliam 1997).
The reasons for the dependence of [Mn/Fe] on metallicity are not
fully understood; possibilities which have been put forward 
include a yield which is sensitive to the neutron excess 
in explosive nucleosynthesis by Type II supernovae
(the odd-even effect), and enhanced Mn production by Type Ia supernovae
(McWilliam 1997 and references therein).
Whatever the reason, our results suggest 
that the nucleosynthetic processes
responsible for the underabundance of Mn at 
${\rm [Fe/H]} < 0$ operate with comparable efficiencies
in the Milky Way and in the galaxies producing DLAs.

\subsection{Silicon}

Silicon if the only $\alpha$~element covered by our observations.
In the ISM of our Galaxy Si is always less depleted than Cr; following 
the same reasoning as above we can use the observed 
[Cr/Zn] ratios to set {\it upper limits} 
to the intrinsic (i.e. corrected for dust depletion)
[Si/Zn] of +0.38, +0.11, and +0.08 in  
Q1351+318 ([Zn/H] = $-0.36$), Q1247+267 ([Zn/H] = $-1.05$),
and Q1354+258 ([Zn/H] = $-1.61$) respectively.
This result is {\it not} in agreement with observations of stars in our 
Galaxy, where [Si/Fe] $\simeq +0.4$ 
at $ -2 \leq {\rm [Fe/H]} \leq -1$ (e.g. Edvardsson 
et al. 1993; McWilliam 1997). 
Thus, on basis of the present data it appears that, 
if the general pattern of dust depletion in the ISM of our Galaxy also 
applies at high $z$, the DLAs observed here do not exhibit the well known 
enhancement of the $\alpha$ elements which is characteristic of the 
metal poor stellar populations of the Milky Way.
Ultimately, this question will only be settled by measuring the ratio of S 
and Zn, two elements which do not suffer from the complications of dust 
depletion and are representative of the $\alpha$ and iron-peak groups 
respectively. 

The relative abundances shown in Figure 9 are broadly similar 
to those reported by Lu et al. (1996, 1998)
for a larger set of DLAs at higher redshifts ($z \simeq 2 - 4$),
but the conclusions reached do differ. 
By focusing mainly on element ratios relative to Fe previous analyses
have generally concluded that the abundances in DLAs are consistent with 
enrichment by Type II supernovae, but have then been faced with the 
conundrum of an inexplicably high Zn abundance.
Unfortunately, dust depletion complicates the interpretation 
of element abundances relative to Fe.
Our approach has been to assume as a starting point that
Zn is undepleted and is a reliable tracer of the iron-peak elements.
The ensuing pattern of element abundances
is consistent with that commonly seen in the ISM of our
Galaxy, reflecting mostly varying degrees of depletions onto dust;
we find no evidence, in our admittedly very limited set of data,  
for an overabundance of the $\alpha$ elements in DLAs of low metallicity.
In his comprehensive analysis of published observations Vladilo (1998)
reached a similar conclusion; we now discuss its implications.\\

\section{DISCUSSION}

The high [$\alpha$/Fe] ratios in metal poor stars are generally thought to 
result from the time delay between the explosions of Type II and Type Ia 
supernovae following a burst of star formation, with the latter producing 
$\sim 2/3$ of the total amount of Fe approximately 1~Gyr after the 
former have enriched the ISM in both $\alpha$ elements and Fe in the 
ratio [$\alpha$/Fe] $\simeq +0.4$\,.
Thus, as emphasized by Gilmore \& Wyse (1991, 1998),
the metallicity at the `turn-over' point in a plot of 
[$\alpha$/Fe] vs. [Fe/H], that is the value of [Fe/H] at which the 
abundance of the $\alpha$ elements decreases from $+0.4$ down to solar, 
is an indication of the past rate of star formation in a galaxy.
In the early chemical evolution of the Milky Way, 
star formation presumably progressed sufficiently fast for
the gas to become enriched to [Fe/H] $\simeq -1$ before 
Type Ia supernovae became important as an additional source of Fe.
But in low surface brightness galaxies or 
in the outer regions of disks, where 
star formation proceeds more slowly 
(McGaugh 1994; Ferguson, Gallagher, \& Wyse 1998),
and in dwarf galaxies, where star formation is often in bursts 
followed by quiescent periods which can last several Gyr (e.g. Grebel 1998),
there may well be sufficient time for the [$\alpha$/Fe] ratio 
to reach
near-solar (or even lower than solar) values while the overall metallicity 
is still below [Fe/H] $ = -1$\,.

In this scenario finding relatively low [Si/Zn] ratios in two metal poor DLAs 
at intermediate redshifts may not be surprising.
Taken together, the results that at $z < 1.5$: 
{\it (a)} the overall metal content of DLAs remains low; {\it (b)} the 
$\alpha$ elements are not enhanced relative to the iron group; and
{\it (c)} often there is no bright galaxy which can be associated with 
the absorber, all point to
the conclusion that {\it current s of intermediate redshift damped \lya\ 
systems do not trace the galaxy population responsible for the bulk of 
star formation at these epochs}.

It seems likely that this is due, at least in part, to the bright 
magnitude limit of {\it HST} observations which makes it difficult 
for metal rich DLAs with large column densities of gas, and 
therefore dust, to be included in existing surveys.
It is as yet unclear to what extent this is also the case at higher 
redshifts, where DLAs are identified from   
ground-based observations.
There are claims of solar [S/Zn] in three systems at \za\ $ > 2$
(Molaro, Centurion, \& Vladilo 1998), 
but a full study has yet to be carried out.

Theoretically there are also reasons to expect that DLAs may 
arise preferentially in galaxies with low rates of star formation, 
irrespectively of any dust obscuration.
Several authors (e.g. Dalcanton, Spergel, \& Summers 1997;
Jimenez et al. 1998; Mao \& Mo 1998; Mo, Mao, \& White 1998)
have emphasized the role which the halo spin parameter plays 
in determining the properties of the galaxies forming within halos of 
cold dark matter. In these models, halos of small angular momentum 
give rise to compact, high density systems, while halos of low mass or 
high angular momentum naturally form disks with a low surface density of 
baryons. For a Schmidt law of star formation (Kennicutt 1998),
the former are the sites of most active star formation,
while the latter dominate the absorption cross-section.
With the limited statistics available at present, it is quite 
possible that QSOs behind the more compact and more metal rich
galaxies have simply not been studied yet.
Thus it appears that, observationally, a large survey of radio selected 
QSOs (where dust obscuration is not an issue) 
is required to ascertain
how damped \lya\ systems fit into the 
broad picture of galaxy formation.

For the moment, the lack of redshift evolution in either the 
neutral gas content (Turnshek 1998) or the metallicity (this work)
of DLAs makes them less useful probes of the star formation history of the 
universe than had been anticipated. The disappointment of this conclusion 
is tempered by the realisation that these metal poor galaxies offer us 
new regimes for testing theories of the nucleosynthesis
of different elements. In closing, we illustrate this point with two topical
examples.

(1) In a recent paper, Kobayashi et al. (1998) have considered the 
effects of metallicity on the evolution of the white dwarf (WD)
progenitors of 
Type Ia supernovae. They make the point that, at 
metallicities below $\sim 1/10$ solar, the optically thick WD wind 
which in the model of Hachisu, Kato, \& Nomoto (1996)
plays a key role in the mass transfer from the binary companion
is reduced to 
the point where the binary system no longer evolves to the supernova stage. 
Measurements of the abundances of S and Zn in DLAs can challenge these 
ideas, if it is found that systems with near solar [$\alpha$/Fe] ratios
are common at low metallicities, as suggested by the work presented here.
In the picture proposed by Kobayashi et al. the $\alpha$ elements are 
expected to remain enhanced until the metal content of the gas from which 
the stars form has increased above [Fe/H] $\sim -1$,  
irrespectively of the timescale of galactic chemical enrichment.
Possibly, the dependence of the WD wind 
on metallicity has been overestimated.

(2) There has been some debate over the last few years as to whether 
[O/Fe] reaches a plateau at $\sim +0.4$ for [Fe/H] $< -1$,
or continues to increase towards lower metallicities. 
The controversy has its origin in the fact that different 
O~I lines are used to measure abundances in dwarf and giant stars,
with conflicting results. Abundances measured from OH lines
in the near-ultraviolet may 
resolve the issue. In a recent study of 23 unevolved halo stars,
Israelian, Garc\'{\i}a L\'{o}pez, \& 
Rebolo (1998) claim that [O/Fe] increases linearly
with decreasing [Fe/H], reaching [O/Fe] $\simeq +1.0$ at [Fe/H] $= -3.0$\,.
The implication is that the Type II supernovae responsible for the 
initial chemical enrichment of the Milky Way synthesized O and Fe in the 
ratio of $\sim10:1$\,.
With a sufficiently large sample of high redshift DLAs it should be 
possible to obtain an independent confirmation of such extreme ratios.

With several new echelle spectrographs soon to be commissioned
on large telescopes we can look forward to a wealth of new 
data of relevance to these and other issues 
concerning  the chemical evolution of galaxies.\\

\acknowledgements

It is a pleasure to acknowledge the competent assistance with the
observations by the staff of the Keck Observatory. Some of the data 
presented here were obtained through the La Palma Service Observations 
program. Our special thanks to Tom Barlow for generously
providing his echelle extraction software, and to Bob Carswell, Jim Lewis 
and Philip Outram for their help at many stages in the data reduction. 
Alice Shapley kindly helped with the Keck observations.
The interpretation of these results benefited from discussions with several
colleagues, particularly J. Prochaska, S. White, and R. Jimenez, 
and from the stimulating environment provided by the
Aspen Center for Physics during a three week workshop in June 1998.
C. C. S. acknowledges support from the National Science
Foundation through grant AST~94-57446 and from the David and Lucile 
Packard Foundation.


%
%

\hspace*{-1cm}
\begin{figure}
\vspace*{-1.5cm}
\figurenum{0}
\epsscale{1.1}
\plotone{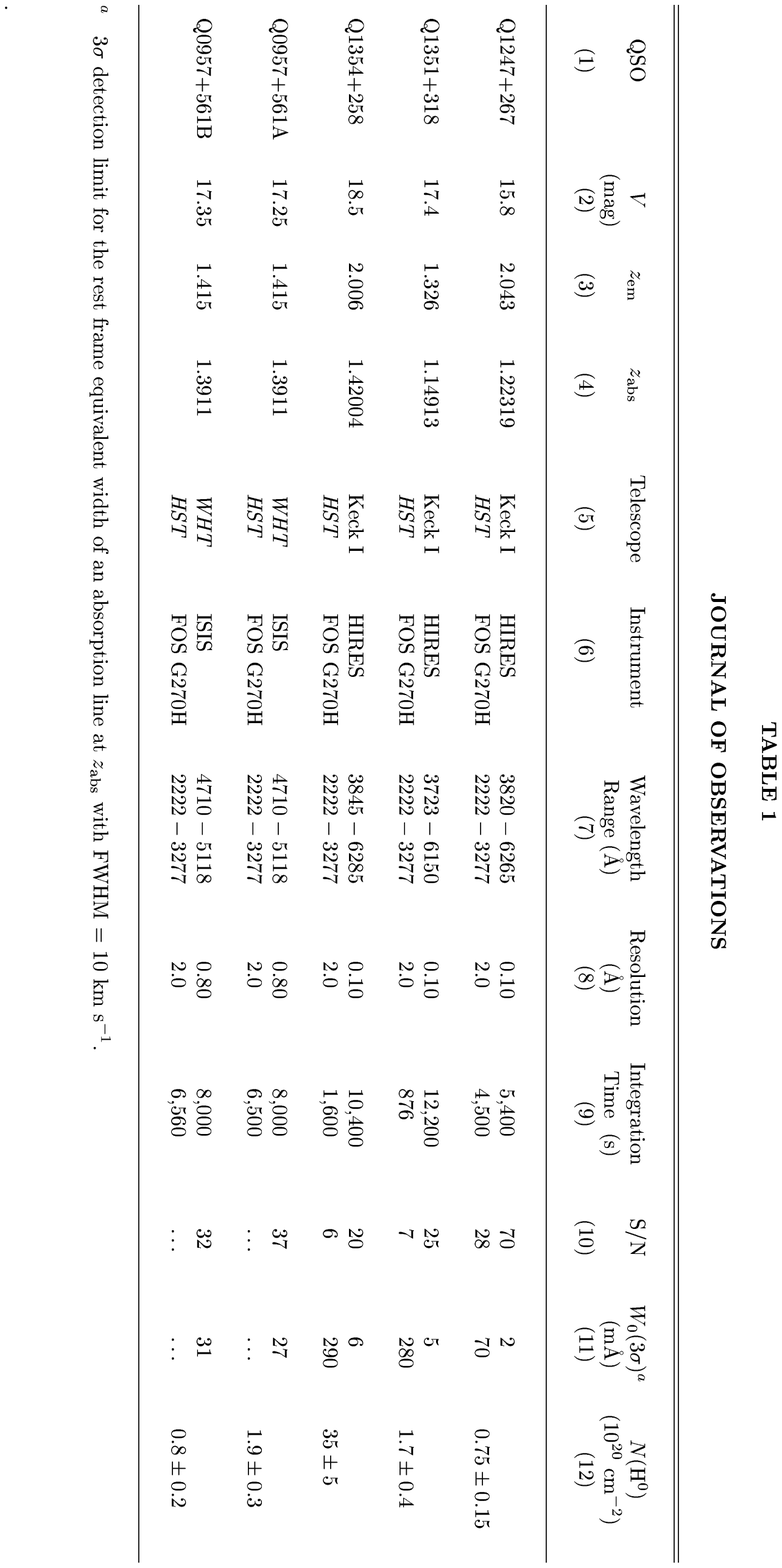}
\end{figure}

%
%

\addtocounter{table}{+1}
\begin{deluxetable}{lllll}
\tablewidth{14cm}
\tablecaption{ABSORPTION LINES IN THE $z_{\rm abs} = 1.22319$ DLA SYSTEM IN Q1247+267}
\tablehead{
\colhead{Line} & \colhead{$\lambda_{\rm obs}^a$ (\AA)} & \colhead{Identification} & \colhead{$z_{\rm abs}^a$} & \colhead{$W_0^b$ (m\AA)} 
}
\startdata
1  & 4019.59  & Si II 1808.0126  & 1.223209 & $12 \pm 2$  \\
2  & 4123.35  & Al III 1854.7164 & 1.223170 & $45 \pm 2$  \\
3  & 4141.31  & Al III 1862.7895 & 1.223177 & $19 \pm 2$  \\
4  & 4504.47  & Zn II 2026.136   & 1.223182 & $6  \pm 1$  \\
5  & 4505.29  & Mg I 2026.4768   & 1.223213 & $5  \pm 1$  \\
6  & 4571.48  & Cr II 2056.254   & 1.223208 & $7  \pm 1$  \\
7  & 4584.73  & Cr II 2062.234   & 1.223186 & $5  \pm 1$  \\
8  & 5001.93  & Fe II 2249.8768  & 1.223202 & $5  \pm 1$  \\
9  & 5026.15  & Fe II 2260.7805  & 1.223192 & $11 \pm 1$  \\
10 & 5211.64  & Fe II 2344.214   & 1.223193 & $160 \pm 1$ \\
11 & 5278.89  & Fe II 2374.4612  & 1.223195 & $96 \pm 1$  \\
12 & 5297.32  & Fe II 2382.765   & 1.223182 & $207 \pm 1$ \\   
13 & 5728.89  & Mn II 2576.877   & 1.223191 & $12  \pm 1$ \\
14 & 5750.62  & Fe II 2586.650   & 1.223192 & $150 \pm 1$ \\
15 & 5768.07  & Mn II 2594.499   & 1.223192 & $8  \pm 1$  \\
16 & 5780.67  & Fe II 2600.1729  & 1.223187 & $218 \pm 1$ \\
17 & 6216.75  & Mg II 2796.352   & 1.223164 & $411 \pm 1$ \\
18 & 6218.51  & Mg II 2796.352   & 1.223794 & $13 \pm 1$  \\
19 & 6232.71  & Mg II 2803.531   & 1.223164 & $354 \pm 1$ \\
20 & 6234.45  & Mg II 2803.531   & 1.223785 & $5 \pm 1$   \\
\enddata
\tablenotetext{a}{Vacuum heliocentric}
\tablenotetext{b}{Rest frame equivalent width and $1 \sigma$ error}
\end{deluxetable}

%
%

\begin{deluxetable}{lllll}
\tablewidth{14cm}
\footnotesize
\tablecaption{ABSORPTION LINES IN THE $z_{\rm abs} = 1.14913$ DLA SYSTEM IN Q1351+318}
\tablehead{
\colhead{Line} & \colhead{$\lambda_{\rm obs}^a$ (\AA)} & \colhead{Identification} & \colhead{$z_{\rm abs}^a$} & \colhead{$W_0^b$ (m\AA)} 
}
\startdata
1  &  3885.68 & Si II 1808.0126  & 1.149144 & $84  \pm 6$  \\
2a &  3986.17 & Al III 1854.7164 & 1.149207 & $219 \pm 5$  \\
2b &  3989.85 & Al III 1854.7164 & 1.151191 & $115 \pm 5$  \\
3a &  4003.38 & Al III 1862.7895 & 1.149132 & $128 \pm 5$  \\
3b &  4007.19 & Al III 1862.7895 & 1.151177 & $71  \pm 5$  \\
4  &  4354.36 & Zn II  2026.136  & 1.149096 & $54  \pm 2$  \\
5  &  4355.11 & Mg I 2026.4768   & 1.149104 & $34  \pm 2$  \\
6  &  4419.20 & Cr II 2056.254   & 1.149151 & $34  \pm 2$  \\
7  &  4431.99 & Cr II 2062.234   & 1.149121 & $22  \pm 3$  \\
8  &  4432.89 & Zn II 2062.664   & 1.149109 & $40  \pm 3$  \\
9  &  4440.44 & Cr II 2066.161   & 1.149126 & $10  \pm 2$  \\
10 &  4835.20 & Fe II 2249.8768  & 1.149095 & $33  \pm 2$  \\
11 &  4858.65 & Fe II 2260.7805  & 1.149103 & $43  \pm 3$  \\
12a & 5038.11 & Fe II 2344.214   & 1.149168 & $393 \pm 3$  \\
12b & 5042.85 & Fe II 2344.214   & 1.151190 & $162 \pm 3$  \\
13a & 5103.04 & Fe II 2374.4612  & 1.149136 & $278 \pm 3$  \\
13b & 5107.93 & Fe II 2374.4612  & 1.151195 & $86  \pm 3$  \\
14a & 5120.97 & Fe II 2382.765   & 1.149171 & $458 \pm 3$  \\   
14b & 5122.65 & Fe II 2382.765   & 1.149876 & $76  \pm 5$  \\
14c & 5125.77 & Fe II 2382.765   & 1.151186 & $229 \pm 4$  \\   
15  & 5538.06 & Mn II 2576.877   & 1.149136 & $88  \pm 4$  \\
16a & 5559.11 & Fe II 2586.650   & 1.149154 & $390 \pm 3$  \\
16b & 5564.30 & Fe II 2586.650   & 1.151161 & $150 \pm 4$  \\
17  & 5575.86 & Mn II 2594.499   & 1.149109 & $53  \pm 2$  \\
18a & 5588.21 & Fe II 2600.1729  & 1.149169 & $501 \pm 3$  \\
18b & 5590.09 & Fe II 2600.1729  & 1.149892 & $79  \pm 5$  \\ 
18c & 5593.46 & Fe II 2600.1729  & 1.151188 & $232 \pm 3$  \\ 
19  & 5601.65 & Mn II 2606.462   & 1.149139 & $50  \pm 4$  \\
20a & 6009.79 & Mg II 2796.352   & 1.149154 & $724 \pm 3$  \\
20b & 6011.84 & Mg II 2796.352   & 1.149887 & $555 \pm 4$  \\
20c & 6014.24 & Mg II 2796.352   & 1.150745 & $33  \pm 3$  \\
20d & 6015.52 & Mg II 2796.352   & 1.151203 & $489 \pm 3$  \\
20e & 6016.63 & Mg II 2796.352   & 1.151600 & $77  \pm 3$  \\
21a & 6025.23 & Mg II 2803.531   & 1.149158 & $668 \pm 3$  \\
21b & 6027.24 & Mg II 2803.531   & 1.149875 & $344 \pm 4$  \\
21c & 6029.42 & Mg II 2803.531   & 1.150652 & $15  \pm 2$  \\
21d & 6030.94 & Mg II 2803.531   & 1.151194 & $416 \pm 3$  \\
21e & 6032.03 & Mg II 2803.531   & 1.151583 & $50  \pm 3$  \\
22a & 6131.48 & Mg I 2852.9642   & 1.149161 & $389 \pm 3$  \\
22b & 6137.22 & Mg I 2852.9642   & 1.151173 & $103 \pm 2$  \\
\enddata                                                  
\tablenotetext{a}{Vacuum heliocentric}
\tablenotetext{b}{Rest frame equivalent width and $1 \sigma$ error}
\end{deluxetable}

%
%

\begin{deluxetable}{llllll}
\tablewidth{17cm}
\tablecaption{ABSORPTION LINES IN THE $z_{\rm abs} = 1.42004$ DLA SYSTEM IN Q1354+258}
\tablehead{
\colhead{Line} & \colhead{$\lambda_{\rm obs}^a$ (\AA)} & \colhead{Identification} & \colhead{$z_{\rm abs}^a$} & \colhead{$W_0^b$ (m\AA)} & \colhead{Comments}
}
\startdata
1  &  3892.50 & Fe II 1608.4545  & 1.420025 & $198 \pm 7$ &\\
2  &  4043.33 & Al II 1670.7874  & 1.420015 & $207 \pm 6$ &\\
3  &  4137.34 & Ni II 1709.600   & 1.420064 & $24  \pm 5$ &\\
4  &  4214.66 & Ni II 1741.549   & 1.420064 & $30  \pm 4$ &\\
5  &  4375.48 & Si II 1808.0126  & 1.420050 & $83  \pm 3$ &\\
6  &  4903.46 & Zn II 2026.136   & 1.420104 & $80  \pm 4$ &Partially blended with Fe~II~$\lambda 2600.17$\\
   &          &                  &          &             &at $z_{\rm abs} = 0.88602$ \\
7  &  4976.25 & Cr II 2056.254   & 1.420056 & $67  \pm 3$ &\\
8  &  4990.72 & Cr II 2062.234   & 1.420055 & $58  \pm 4$ &\\
9  &  4991.82 & Zn II 2062.664   & 1.420084 & 60:         &Absorption line abnormally broad \\
10 &  5000.21 & Cr II 2066.161   & 1.420049 & $39  \pm 2$ &\\
11 &  5444.85 & Fe II 2249.8768  & 1.420065 & $62  \pm 4$ &\\
12 &  5471.19 & Fe II 2260.7805  & 1.420045 & $84  \pm 3$ &\\
13 &  5673.04 & Fe II 2344.214   & 1.420018 & $326 \pm 3$ &\\
14 &  5746.26 & Fe II 2374.4612  & 1.420027 & $260 \pm 3$ &\\
15 &  5766.34 & Fe II 2382.765   & 1.420020 & $371 \pm 3$ &\\   
16 &  6236.21 & Mn II 2576.877   & 1.420065 & $96  \pm 3$ &\\
17 &  6259.75:& Fe II 2586.650   & 1.420022:& 375:        &Affected by a cosmic ray \\
18 &  6278.77 & Mn II 2594.499   & 1.420032 & $73  \pm 4$ & \\
\enddata
\tablenotetext{a}{Vacuum heliocentric}
\tablenotetext{b}{Rest frame equivalent width and $1 \sigma$ error}
\end{deluxetable}

%
%

\begin{deluxetable}{ccccc}
\tablewidth{12cm}
\tablecaption{COMPONENT STRUCTURE IN THE $z_{\rm abs} = 1.22319$ DLA SYSTEM IN Q1247+267}
\tablehead{
\colhead{Component No.} & \colhead{$z_{\rm abs}$} 
& \colhead{$b$ (km~s$^{-1}$)} & \colhead{log $N$\/(Fe$^+$) (cm$^{-2}$)}
}
\startdata
1 & 1.223102 & 5.4 & 12.64 \\
2 & 1.223202 & 6.3 & 13.95 \\
\enddata
\end{deluxetable}

%
%

\begin{deluxetable}{ccccc}
\tablewidth{12cm}
\tablecaption{COMPONENT STRUCTURE IN THE $z_{\rm abs} = 1.14913$ 
DLA SYSTEM IN Q1351+318}
\tablehead{
\colhead{Component No.} & \colhead{$z_{\rm abs}$} 
& \colhead{$b$ (km~s$^{-1}$)} & \colhead{log $N$\/(Fe$^+$) (cm$^{-2}$)}
}
\startdata
1 & 1.148895 & 5.3  & 12.23 \\
2 & 1.149033 & 3.5  & 14.01 \\
3 & 1.149139 & 10.1 & 14.50 \\
4 & 1.149318 & 7.7  & 13.40 \\
5 & 1.149599 & 3.3  & 11.86 \\
6 & 1.149778 & 23.3 & 12.27 \\
7 & 1.150051 & 14.1 & 12.42 \\
8 & 1.150231 & 8.2  & 11.81 \\
9 & 1.151012 & 3.5  & 11.89 \\
10 & 1.151158 & 7.4  & 13.60 \\
11 & 1.151177 & 3.0  & 13.68 \\
12 & 1.151331 & 4.9  & 12.63 \\
13 & 1.151519 & 3.9  & 11.69 \\
\enddata
\end{deluxetable}

%
%

\begin{deluxetable}{ccccc}
\tablewidth{12cm}
\tablecaption{COMPONENT STRUCTURE IN THE $z_{\rm abs} = 1.42004$ DLA SYSTEM IN Q1354+258}
\tablehead{
\colhead{Component No.} & \colhead{$z_{\rm abs}$} 
& \colhead{$b$ (km~s$^{-1}$)} & \colhead{log $N$\/(Fe$^+$) (cm$^{-2}$)}
}
\startdata
1 & 1.419879 & 3.5 & 13.52 \\
2 & 1.420053 & 8.1 & 15.02 \\
\enddata
\end{deluxetable}

%
%
\begin{figure}
\vspace*{-1.5cm}
\figurenum{0}
\epsscale{1.1}
\plotone{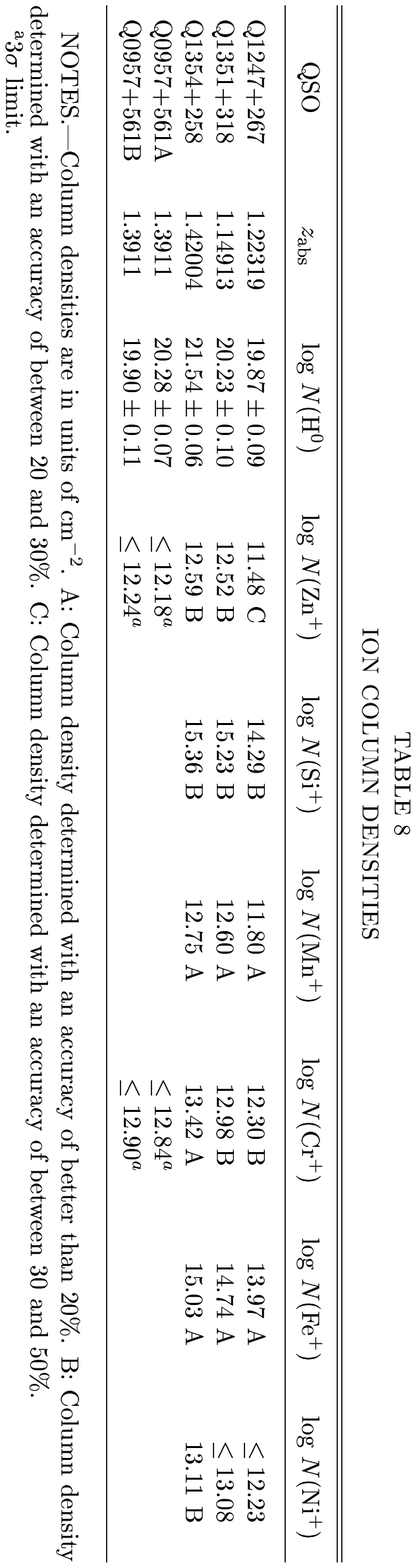}
\end{figure}

%
%
\hspace*{-1cm}
\begin{figure}
\vspace*{-1.5cm}
\figurenum{0}
\epsscale{1.1}
\plotone{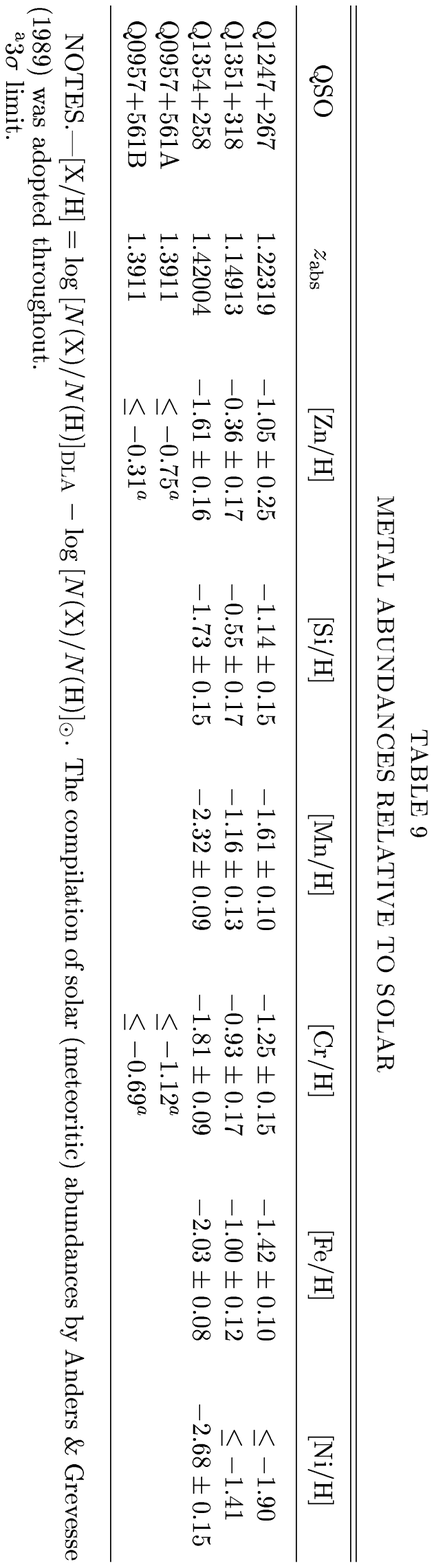}
\end{figure}


%
%

\addtocounter{table}{+2}
\begin{deluxetable}{lcccccc}
\tablewidth{0pc}
\scriptsize
\tablecaption{COLUMN DENSITY WEIGHTED METALLICITIES}
\tablehead{
\colhead{} & \colhead{Redshift Range} & \colhead{Lookback Time (Gyr)$^a$} &
\colhead{DLAs} & \colhead{Detections} &
\colhead{Upper Limits} 
& \colhead{${\rm [}\langle{\rm Zn/H}_{\rm DLA}\rangle{\rm ]}$} 
}
\startdata
1 &  0.40 $-$ 1.49 & ~5.4 $-$ 11.4 &10  &  9 &  1  & $-1.03 \pm 0.23$ \nl
2 &  1.50 $-$ 1.99 & 11.4 $-$ 12.7 & 8  &  6 &  2  & $-0.96 \pm 0.17$ \nl
3 &  2.00 $-$ 2.49 & 12.7 $-$ 13.6 &12  &  6 &  6  & $-1.23 \pm 0.11$ \nl
4 &  2.50 $-$ 2.99 & 13.6 $-$ 14.3 & 7  &  3 &  4  & $-1.11 \pm 0.09$ \nl
5 &  3.00 $-$ 3.49 & 14.3 $-$ 14.8 & 3  &  0 &  3  & $\leq -1.39$ \nl
\enddata
\tablenotetext{a}{$H_0 = 50$~km~s$^{-1}$~Mpc$^{-1}$; $q_0 = 0.01$}
\end{deluxetable}

\vfill

%
%

%
%

\newpage
\begin{figure}
\figurenum{1}
\epsscale{1.1}
\hspace{-1.7cm}
\plotone{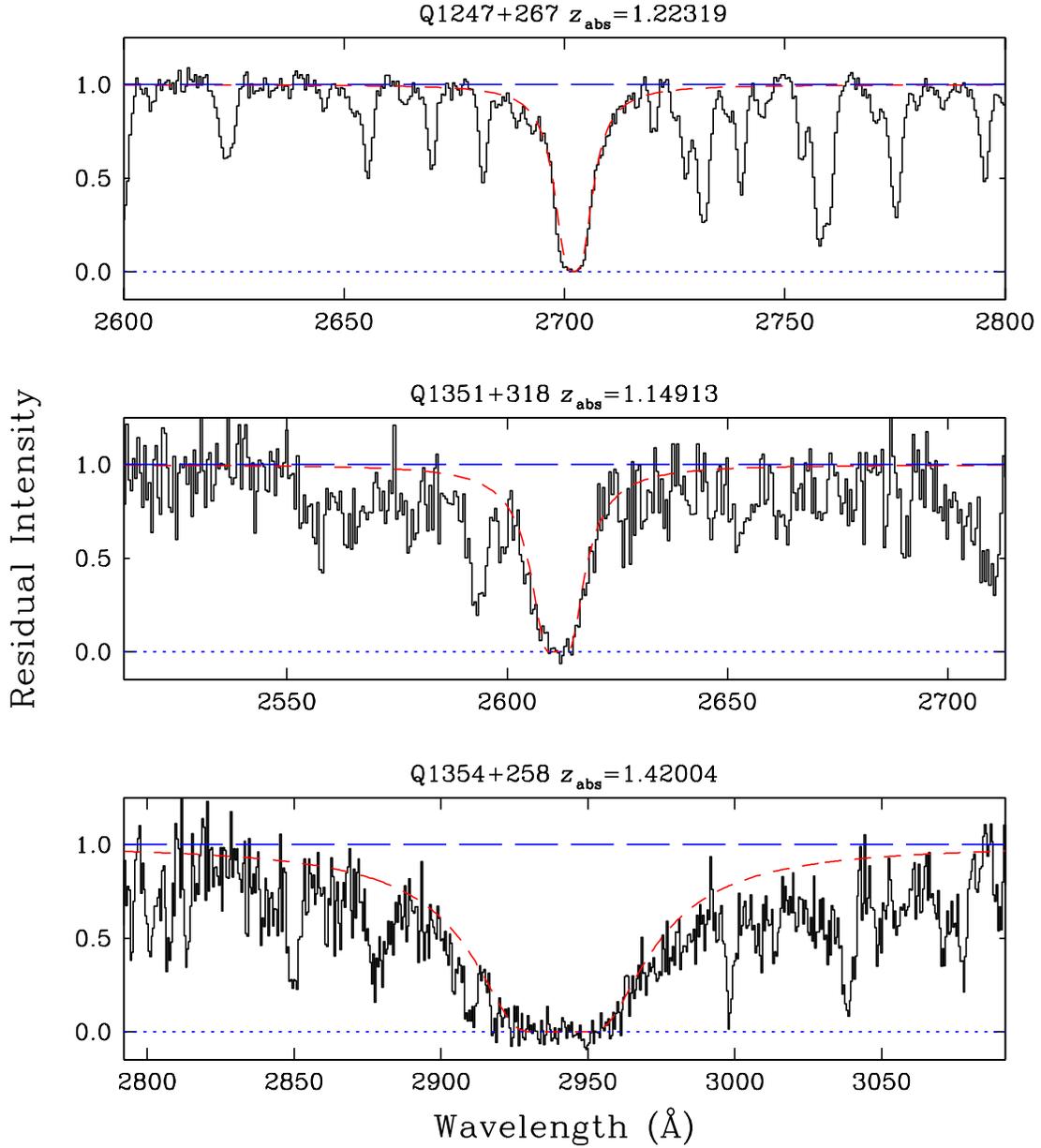}
\vspace{-5cm}
\caption{Portions of the three FOS spectra retrieved from the {\it
HST} data archive; details of the observations are collected in Table
1. The spectra have been normalised to the QSO continua. The
short-dash lines show the damped \lya\ absorption profiles
corresponding to the values of \nhi\ listed in column (12) of Table
1.}
\end{figure}

%
%

\newpage
\begin{figure}
\figurenum{2}
\epsscale{1.1}
\hspace{-1.8cm}
\plotone{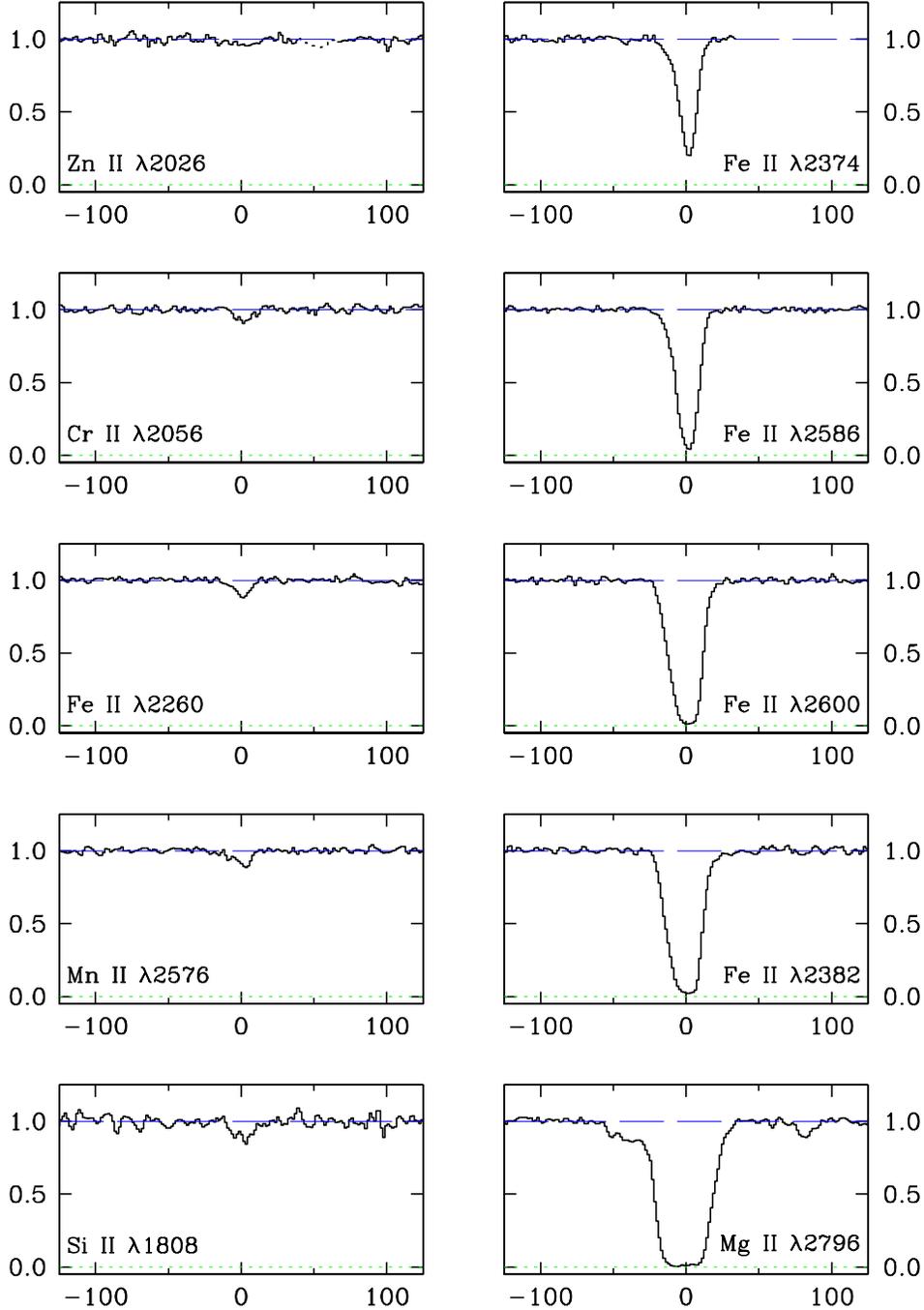}
\vspace{-3.5cm}
\caption{Profiles of selected absorption lines in the 
$z_{\rm abs} = 1.22319$ DLA in Q1247+267. 
The $y$-axis is residual intensity and the $x$-axis is relative
velocity in \kms.}
\end{figure}

%
%

\newpage
\begin{figure}
\figurenum{3}
\epsscale{1.1}
\hspace{-1.8cm}
\plotone{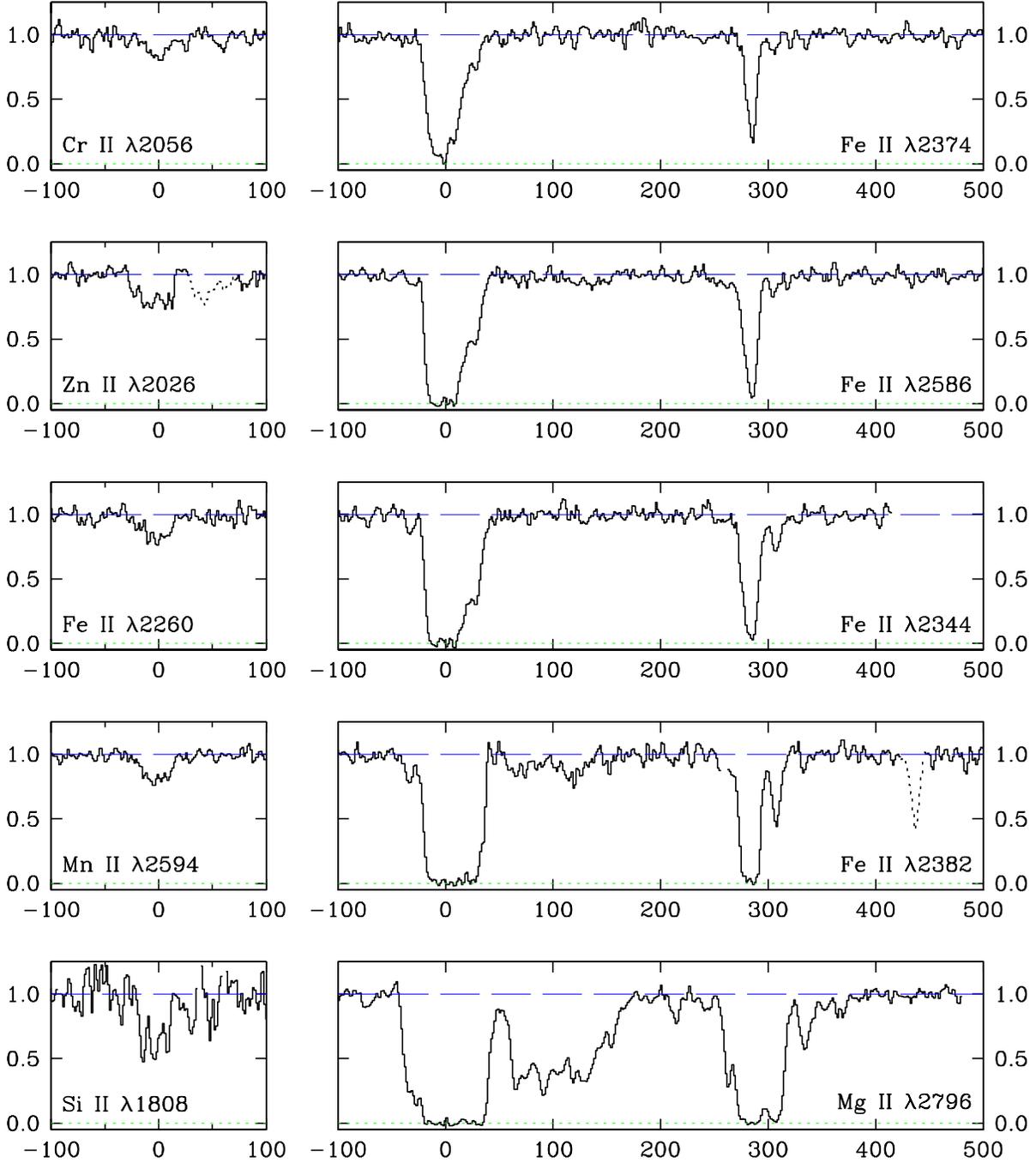}
\vspace{-3.5cm}
\caption{Profiles of selected absorption lines in the 
$z_{\rm abs} = 1.14913$ DLA in Q1351+318. 
The $y$-axis is residual intensity and the $x$-axis is relative
velocity in \kms.}
\end{figure}

%
%

\newpage
\begin{figure}
\figurenum{4}
\epsscale{1.1}
\hspace{-1.8cm}
\plotone{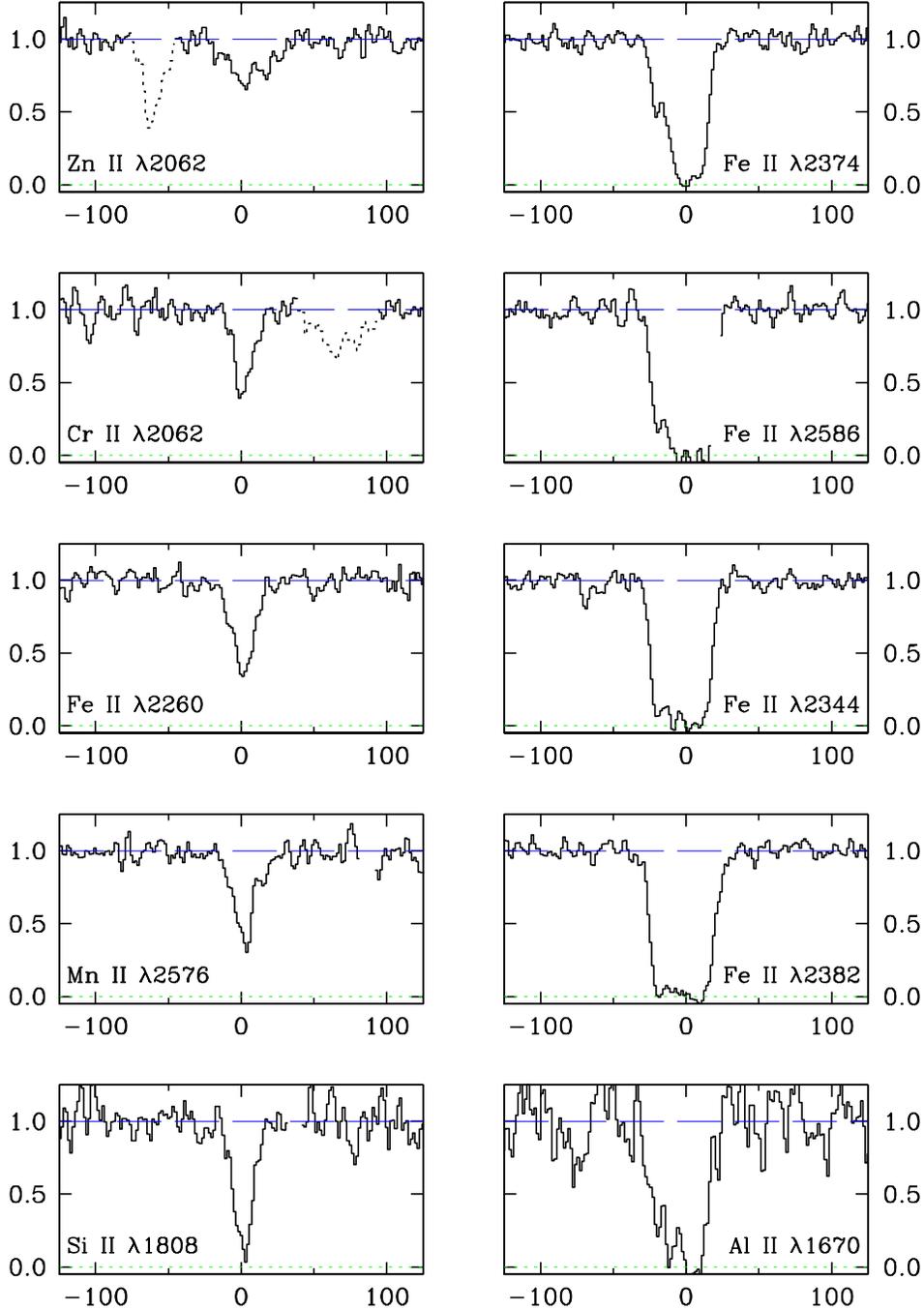}
\vspace{-3.5cm}
\caption{Profiles of selected absorption lines in the 
$z_{\rm abs} = 1.42004$ DLA in Q1354+258. 
The $y$-axis is residual intensity and the $x$-axis is relative
velocity in \kms. The red wing of Fe~II~$\lambda 2586$ is affected by
a cosmic-ray.}
\end{figure}

%
%

\newpage
\begin{figure}
\figurenum{5}
\epsscale{1.1}
\hspace{-1.7cm}
\plotone{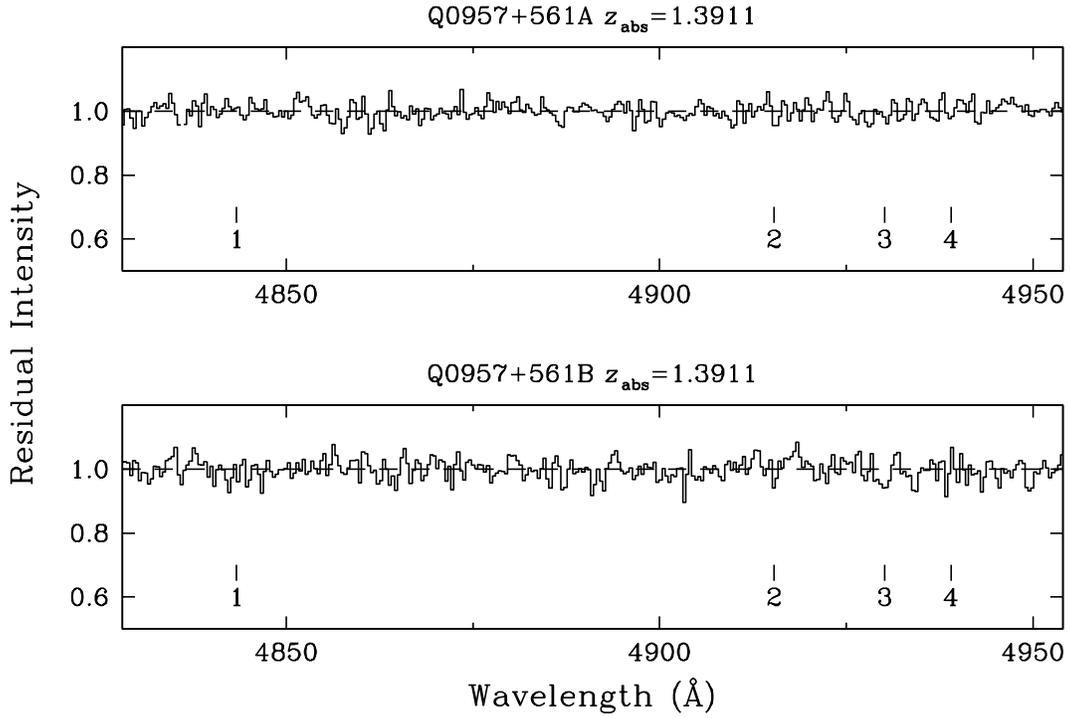}
\vspace{-8cm}
\caption{
Portions of the WHT spectra of Q0957+561A, B.
The vertical tick  marks  indicate
the expected positions of absorption lines 
in the \za\ = 1.3911 DLA.
Line 1:~Zn~II~$\lambda 2026.136$; 
line 2:~Cr~II~$\lambda 2056.254$; 
line 3:~Cr~II~$\lambda   2062.234$~+~Zn~II~$\lambda  2062.664$  (blended);  
and  line 4:~Cr~II~$\lambda 2066.161$.  
The spectra have been normalised to the
underlying continua and are shown on an expanded vertical scale.
} 
\end{figure}

%
%

\newpage
\begin{figure}
\figurenum{6}
\epsscale{1.1}
\hspace{-1.8cm}
\plotone{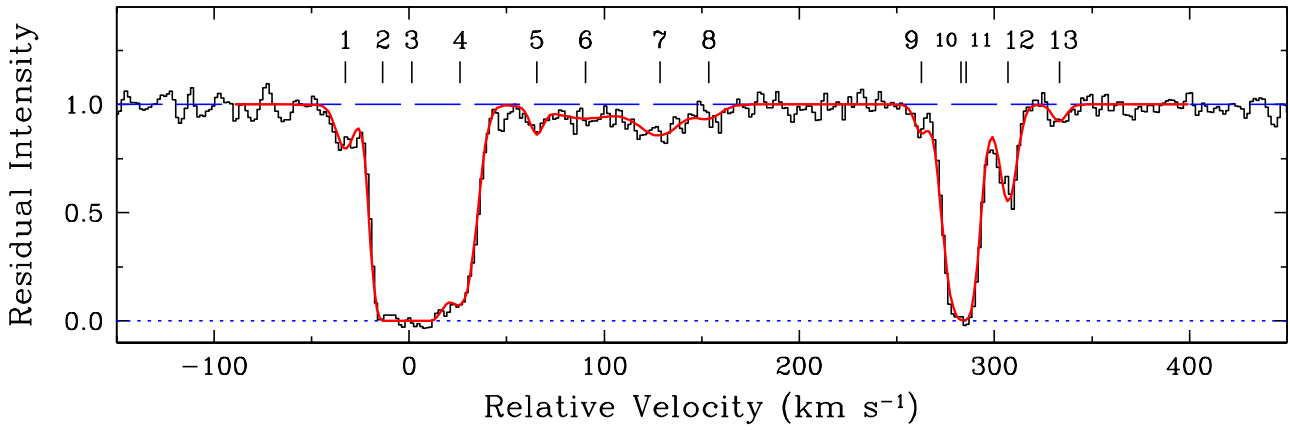}
\vspace{-14cm}
\caption{Profile fit to the Fe~II~$\lambda 2600.1729$ line in the
\za\ = 1.14913 DLA in Q1351+318. Observed and computed 
profiles are shown by the histogram and the continuous line
respectively. Vertical tick marks indicate the positions of
individual absorption components.}
\end{figure}

%
%

\newpage
\begin{figure}
\figurenum{7}
\hspace{-1.8cm}
\plotone{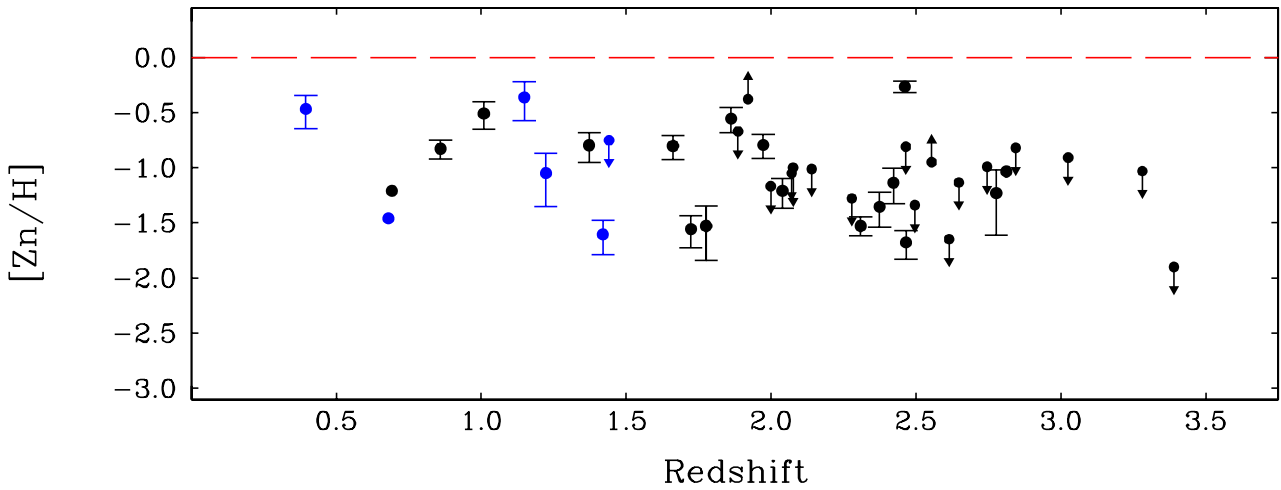}
\vspace{-14cm}
\caption{Plot of the abundance of Zn against redshift for the
full sample of DLAs, consisting of the 34 systems in the study by
Pettini et al. (1997a) plus the six new measurements at $z <
1.5$ considered
here. Abundances are measured on a log scale relative to the
solar value shown by the broken line at [Zn/H] = 0.0\,. Upper
limits, corresponding to  non-detection  of  the  Zn~II lines,
are indicated by downward-pointing arrows.}
\end{figure}

%
%

\newpage
\begin{figure}
\figurenum{8}
\epsscale{1.1}
\hspace{-1.8cm}
\plotone{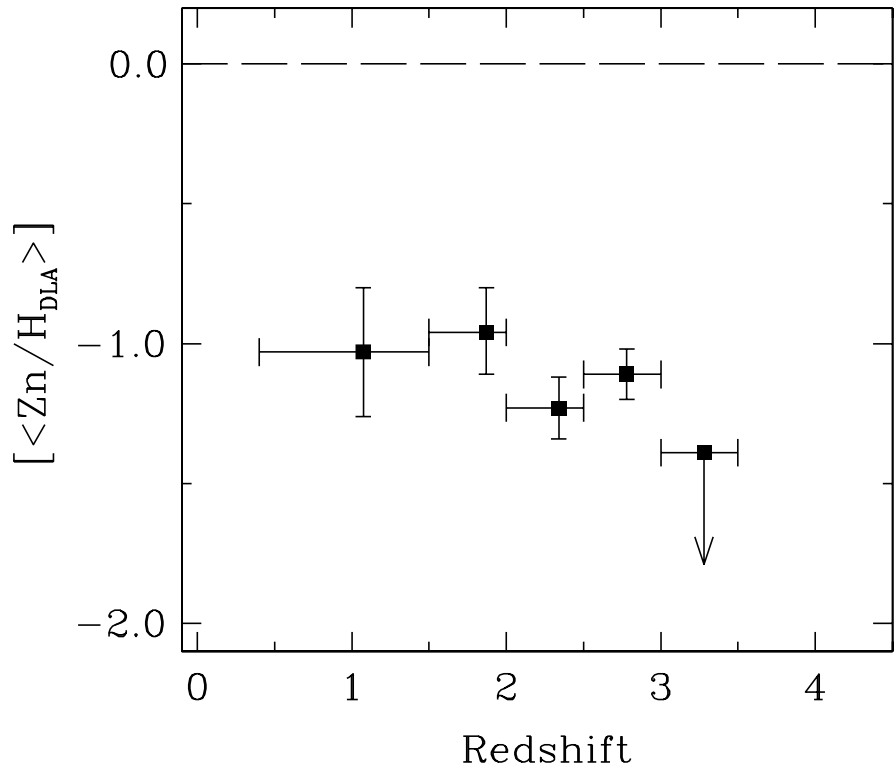}
\vspace{-12cm}
\caption{Column density-weighted metallicities for the full sample of
DLAs. The symbols have been plotted at the median redshift of the
DLAs in each bin.}
\end{figure}

%
%

\newpage
\begin{figure}
\figurenum{9}
\epsscale{0.95}
\hspace{-1.8cm}
\plotone{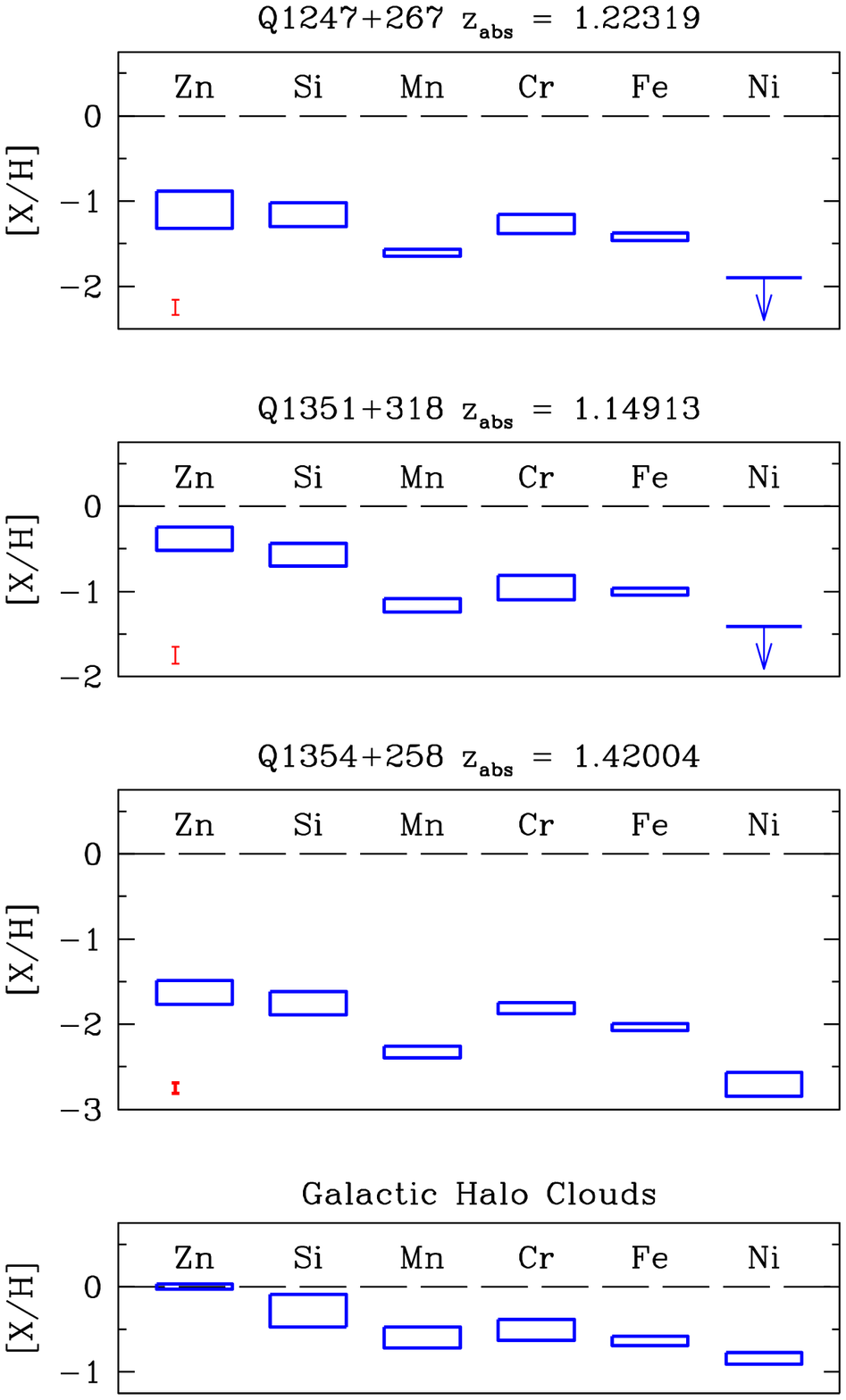}
\vspace{-3.5cm}
\caption{Element abundances in the three damped \lya\ systems studied 
are plotted on a logarithmic scale relative to solar values. The height of 
each box represents the uncertainty in 
the column density of that element; the 
vertical bars near the bottom left hand corners of the panels 
indicate the errors 
in the column densities of neutral hydrogen. The bottom panel shows the 
abundances of the same elements relative to Zn (assumed to be solar)
in local interstellar clouds located out of the plane of the Galaxy, where
Si, Mn, Cr, Fe, and Ni are partly depleted onto dust (reproduced from the 
compilation by Savage \& Sembach 1996).
}
\end{figure}


\newpage

\begin{references}
\reference{} Anders, E., \& Grevesse, N. 1989, Geochim.
Cosmochim. Acta, 53, 197

\reference{} Blades, J.C., Wheatley, J.M., Panagia, N., Grewing,
M., Pettini, M., \& Wamsteker, W. 1988, ApJ, 334, 308.

\reference{} Boiss\'{e}, P., Le Brun, V., Bergeron, J., \& Deharveng, J.
1998, A\&A, 333, 841

\reference{} Bowen, D.V., Roth, K.C., Blades, J.C., \& Meyer,
D.M. 1994, ApJ, 420, L71

\reference{} Dalcanton, J.J., Spergel, D.N., \& Summers, F.J. 1997,
ApJ, 482, 659

\reference{} de la Varga, A., \& Reimers, D. 1998, in Structure
and Evolution of the Intergalactic Medium from QSO Absorption
Line Systems, ed. P. Petitjean, \& S. Charlot (Paris: Editions
Frontieres),

\reference{} Edvardsson, B., Andersen, J., Gustafsson, B.,
Lambert, D.L., Nissen, P.E., \& Tomkin, J. 1993,  A\&A, 275, 101 

\reference{} Efron, B., \& Tibshirani, R.J. 1993, An Introduction to the 
Bootstrap, (New York: Chapman \& Hall)

\reference{} Fall, S.M., \& Pei, Y.C. 1993, ApJ, 402, 479

\reference{} Ferguson, A.M.N., Gallagher, J.S., \& Wyse, R.F.G. 1998, 
ApJ, in press (astro-ph/9805166)

\reference{} Gardner, J.P., Katz, N., Weinberg,, D.H., \& Hernquist, L.
1997, ApJ, 486, 42

\reference{} Gilmore, G., \& Wyse, R.F.G. 1991, ApJ, 367, L55

\reference{} Gilmore, G., \& Wyse, R.F.G. 1998, AJ, in press 
(astro-ph/9805144)

\reference{} Grebel, E.K. 1998, in Dwarf Galaxies and Cosmology, ed.
Thuan, T.X., Balkowski, C., Cayatte, V., \& Van, T.T. 
(Paris: Editions Frontieres), in press (astro-ph/9806191)

\reference{} Hachisu, I., Kato, M., \& Nomoto, K. 1996, ApJ, 470, L97

\reference{} Haehnelt, M.G., Steinmetz, M., \& Rauch, M. 1998, ApJ,
495, 647

\reference{} Israelian, G., Garc\'{\i}a L\'{o}pez, R.J., \& Rebolo, R. 1998,
(astro-ph/9806235)

\reference{} Jannuzi, B.T., et al. 1998, ApJS, 118, in press
(astro-ph/9805148)

\reference{} Jimenez, R., Padoan, P., Matteucci, F., \& Heavens, A.F.
1998, MNRAS, 299, 123.

\reference{} Kennicutt, R. 1998, ARA\&A, in press

\reference{} Kobayashi, C., Tsujimoto, T., Nomoto K., Hachisu, I., 
\& Kato, M. 1998, ApJ, in press (astro-ph/9806335)

\reference{} Kulkarni, V.P., \& Fall, S.M. 1993, ApJ, 413, L63

\reference{} Lanzetta, K.M., et al. 1997, AJ, 114, 1337 

\reference{} Lanzetta, K.M., Wolfe, A.M., \& Turnshek, D.A. 1995,
ApJ, 440, 435

\reference{} Le Brun, V. Bergeron, J., Boiss\'{e}, P., \& Deharveng, J.M.
1997, A\&A, 321, 733

\reference{} Lu, L., Sargent, W.L.W., Barlow, T.A., Churchill, C.W.,
\& Vogt, S.S. 1996, ApJS, 107, 475

\reference{} Lu, L., Sargent, W.L.W., \& Barlow, T.A. 1998,
in Cosmic Chemical Evolution, (Kluwer Academic Publishers),
in press (astro-ph/9710370) 

\reference{} Madau, P., Pozzetti, L., \& Dickinson, M.E. 1998,
ApJ, 498, 106


\reference{} Mao, S., \& Mo, H.J. 1998, MNRAS, submitted (astro-ph/9805094)

\reference{} Mo, H.J., Mao, S., \& White, S.D.M. 1998, MNRAS, submitted
(astro-ph/9807341)

\reference{} McGaugh, S.S. 1994, ApJ, 426, 135

\reference{} McWilliam, A. 1997, ARA\&A, 35, 503. 

\reference{} McWilliam, A., Preston, G.W., Sneden, C., \& Searle, L. 
1995, AJ, 109, 2757

\reference{} Michalitsianos, A.G. et al. 1997, ApJ, 474, 598

\reference{} Molaro, P., Centurion, M., \& Vladilo, G. 1998, MNRAS, 293, 37

\reference{} Morton, D.C. 1991, ApJS, 77, 119

\reference{} Peacock, J.A., Jimenez, R., Dunlop, J.S., Waddington, I.,
Spinrad, H., Stern, D., Dey, A., \& Windhorst, R.A. 1998, MNRAS, 296,
1089

\reference{} Pei, Y.C., \& Fall, S.M. 1995, ApJ, 454, 69

\reference{} Pettini, M. \& Bowen, D.V. 1997, A\&A, 327, 22

\reference{} Pettini, M., King, D.L., Smith, L.J., \& Hunstead, R.W.
1997b, ApJ, 478, 536

\reference{} Pettini, M., Smith, L.J., King, D.L., \& Hunstead, R.W.
1997a, ApJ, 486, 665

\reference{} Prochaska, J.X., \& Wolfe, A.M. 1996, ApJ, 470, 403 

\reference{} Prochaska, J.X., \& Wolfe, A.M. 1997, ApJ, 487, 73 

\reference{} Rao, S.M., \& Turnshek, D.A. 1998, ApJ, in press
(astro-ph/9805093)

\reference{} Ryan, S.G., Norris, J.E., \& Beers, T.C. 1996, ApJ, 471, 254

\reference{} Savage, B.D., \& Sembach, K.R. 1996, ARA\&A, 34, 279

\reference{} Sneden, C., Gratton, R.G., \& Crocker, D.A. 1991, A\&A, 246, 
354

\reference{} Steidel, C.C., Pettini, M.,  Dickinson, M., \&.Persson,
S.E. 1994, AJ, 108, 2046

\reference{} Storrie-Lombardi, L., McMahon, R.G., \& Irwin, M.J. 1996,
MNRAS, 283, L79


\reference{} Turnshek, D.A. 1998, in Structure and Evolution of the
Intergalactic Medium from QSO Absorption Line Systems, ed. P.
Petitjean, \& S. Charlot (Paris: Editions Frontieres), 263
                                  
\reference{} Turnshek, D.A., \& Bohlin, R.C. 1993, ApJ, 407, 60

\reference{} Viegas, S.M. 1995, MNRAS, 276, 268

\reference{} Vladilo, G. 1998, ApJ, 493, 583

\reference{} Vogt, S.S. et al. 1994, S.P.I.E., 2198, 362

\reference{} Welty, D.E., Lauroesch, J.T., Blades, J.C., 
Hobbs, L.M., \& York, D.G. 1997, ApJ, 489, 672

\reference{} Wolfe, A.M., \& Prochaska, J.X. 1998, ApJ, 494, L15

\reference{} Wolfe, A.M., Turnshek, D.A., Smith, H.E., \& Cohen, R.D. 
1986, ApJS, 61, 249

\reference{} Zuo, L., Beaver, E.A., Burbidge, E.M., Cohen, R.D.,
Junkkarinen, V.T., \& Lyons, R.W. 1997, ApJ, 477, 568

\reference{} Zwaan, M. 1998, in Dwarf Galaxies and Cosmology, ed.
Thuan, T.X., Balkowski, C., Cayatte, V., \& Van, T.T. 
(Paris: Editions Frontieres), in press (astro-ph/9806260)
\end{references}
\end{document}